\pdfoutput=1
\documentclass[conference]{IEEEtran}
\IEEEoverridecommandlockouts % Needed for the license information in the footnote
\usepackage{cite}
\usepackage{amsmath,amssymb,amsfonts}
\usepackage{graphicx}
\usepackage{textcomp}
\usepackage{xcolor}
\usepackage{algorithm}
\usepackage{algpseudocode}
\def\BibTeX{{\rm B\kern-.05em{\sc i\kern-.025em b}\kern-.08em
		T\kern-.1667em\lower.7ex\hbox{E}\kern-.125emX}}
\usepackage{hyperref}

\begin{document}

\title{Enso\\
	\large A general-purpose virtual machine
	\thanks{This work is licensed under the Creative Commons Attribution-ShareAlike 4.0 International License.}
}

\author{
	\IEEEauthorblockN{Bruno Fran\c{c}a}
	\IEEEauthorblockA{
		Trinkler Software\\
		\textit{company@trinkler.software}}
	\\\\Version 4 -- August 1, 2019
}

\maketitle
\thispagestyle{plain} %to have page numbers
\pagestyle{plain} %to have page numbers

\begin{abstract}
In this paper we introduce Enso, a virtual machine designed to be used as general-purpose state transition function in blockchains. This design allows the blockchain application logic to be coded into the state, instead of into the state transition function, making it much more flexible and easier to modify. A byproduct is reducing the frequency of forks, concerted or not.
\end{abstract}

\section{Introduction}
Despite the diversity and complexity in the blockchain world today, all blockchains are just distributed virtual machines.

Independently of the function for which it is used, a blockchain is just a virtual machine running some code distributed across several computers.

The components of such a distributed virtual machine can be visualized as a stack:

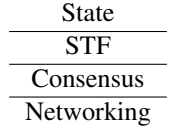
\begin{center}
	\begin{tabular}{c}
		State\\
		\hline
		STF\\
		\hline
		Consensus\\
		\hline
		Networking
	\end{tabular}
\end{center}

Starting at the bottom, we have the networking. This basically includes all the tools that the computer needs to connect to the other nodes and to receive, transmit and relay data across the network. It is a standard component for any peer-to-peer network.

Then we have the consensus. This allows all nodes to reach an agreement on what blocks are accepted. There is a huge variety of consensus protocols, but they all have the same objective: making all honest nodes reach an agreement on some input data.

Lastly, we have the state and the state transition function (STF). State represents, of course, the current state of the virtual machine. The state, for most users, is the only component that they interact with. The state transition function is the function that takes as input the previous state and a block, and produces a new state as output. So, it is what defines the rules for modifying the state. Contrary to the networking and consensus, these two components no longer interact with, or have any notion of, other nodes. Together they just form a simple virtual machine.

The way in which each of these components is modified, or upgraded, varies from blockchain to blockchain. The following image illustrates how legacy chains, like Bitcoin \cite{nakamoto2008bitcoin} or Ethereum \cite{buterin2014next}, are modified:

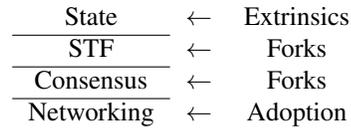
\begin{center}
	\begin{tabular}{c c c}
		State & $\leftarrow$ & Extrinsics\\
		\cline{1-1}
		STF & $\leftarrow$ & Forks\\
		\cline{1-1}
		Consensus & $\leftarrow$ & Forks\\
		\cline{1-1}
		Networking & $\leftarrow$ & Adoption
	\end{tabular}
\end{center}

The state is always modified with extrinsics\footnote{We use the same definition for transaction as the one taken by Parity. There are extrinsics, which are any input to the state transition function, and both transactions and inherents are mutually exclusive types of extrinsics. Transactions are extrinsics that are propagated through the network and signed. Inherents are neither propagated nor signed. An example of an inherent is a timestamp.}. This is how most people will interact with the blockchain and it is very straightforward.

Next we have the state transition function and the consensus. Both can be modified with forks, either soft forks or hard forks. The problem with forking is that forks are difficult to organize and can cause the chain to split into two networks (ex: Bitcoin and Bitcoin Cash, or Ethereum and Ethereum Classic).

Finally we have the network. Changes in the network normally are not contentious since they can happen with gradual user adoption. For example, some nodes in the Bitcoin network can start accepting transactions via Tor and relaying them to miners. There is no need for any type of consensus on whether transactions should be accepted via Tor. Any node that wishes to do so can make that change and make that service available to the network right away. So, networking updates happen with organic user adoption.

Recently, blockchains appeared which claim to be upgradable by \textit{governance}. We are only interested in the technical aspect of upgrading a blockchain and not in the social or political aspect (e.g. if voting should be proportional to stake, etc). The most famous examples in this category are probably Tezos \cite{goodman2014tezos} and Parity's Substrate (Substrate is not actually a blockchain but rather a blockchain development kit). With this type of chain we have the following stack:

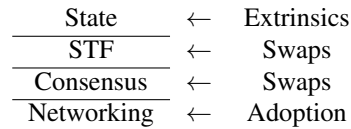
\begin{center}
	\begin{tabular}{c c c}
		State & $\leftarrow$ & Extrinsics\\
		\cline{1-1}
		STF & $\leftarrow$ & Swaps\\
		\cline{1-1}
		Consensus & $\leftarrow$ & Swaps\\
		\cline{1-1}
		Networking & $\leftarrow$ & Adoption
	\end{tabular}
\end{center}

Note that the state and networking are still modified in the same fashion as legacy chains, the difference is that the state transition function (STF) and the consensus use \textit{module swaps} instead of forks.

In these blockchains, there is a shell that contains only the networking and some other auxiliary code. Then there are two slots (one for the STF, another for the consensus) that accept arbitrary code blocks, called \textit{modules}.

The idea is that a blockchain can be modified by simply swapping one module for another. If the blockchain also has governance functionality coded into it, it is possible for the nodes to reach an agreement on whether a module should be swapped or not.

Governance and module swaps are different components. Governance only allows a network to reach an agreement on whether the blockchain is to be updated or not. The actual method of doing that update is the module swapping. Objectively, the main difference between forks and swaps is that swaps are concerted, while forks are not. Otherwise the process is similar. The miners (or validator, or authorities) all agree to, at a specific time, stop using one module and start using the next one.

The breakthrough on the part of Tezos was realizing that legacy blockchains have their governance \textit{off-chain} and that instead the governance should be placed \textit{on-chain} into the STF.

Another way of analyzing this stack is seeing what functionality is coded into each component. As we have said, a blockchain is a distributed virtual machine, but of course, each blockchain is running a specific application (also commonly called the \textit{runtime}). For example, Bitcoin is a distributed virtual machine running a ledger application. We can visualize where each part of the application is located in the stack:

\begin{center}
	\begin{tabular}{c c c}
		App data & $\equiv$ & State\\
		\cline{1-1} \cline{3-3}
		App logic & $\equiv$ & STF\\
		\cline{3-3}
		 &  & Consensus\\
		\cline{3-3}
		&  & Networking
	\end{tabular}
\end{center}

The application logic and data are separated, with the logic being hardcoded into the STF and the data being stored into the state.

This separation of \textit{code} and \textit{data} is how early computers worked. Programs were hardcoded into the hardware using plugboards and then data would be fed into the computer using some storage medium (for example, punch cards). This was time-consuming, requiring the plugboard to be rewired whenever a new program was to be run in the computer. Then along came the stored-program computer, which treats code and data equally. The paradigm change was thinking of the hardware as a general-purpose shell without any specific functionality predefined into it. Now programs could be fed into the computer just like regular data.

Enso applies the stored-program computer concept to blockchains. It adopts a very simple general-purpose virtual machine as the state transition function so that the application logic can be put into the state. This allows for both the application logic and data to be modified with extrinsics, while the consensus and the virtual machine itself are still modified with forks or module swaps.

\begin{center}
	\begin{tabular}{c c c c c}
		App logic and data & $\equiv$ & State & $\leftarrow$ & Extrinsics\\
		\cline{1-1} \cline{3-3}
		Virtual machine & $\equiv$ & STF & $\leftarrow$ & Forks/Swaps\\
		\cline{3-3}
		 &  & Consensus & $\leftarrow$ & Forks/Swaps\\
		\cline{3-3}
		 & & Networking & $\leftarrow$ & Adoption
	\end{tabular}
\end{center}

By treating the logic and the data equally, blockchains can be built with a greater degree of flexibility. And extrinsics allow for simpler and more granular modifications to the application logic. Changes to the consensus or to the virtual machine still need to be done using the module swaps, but we expect that consensus and virtual machine updates will be fairly rare, at least when compared with changes to the application logic.

\section{Actor model}
The actor model is a mathematical model of computation, especially useful when concurrency is desired, that was first published in 1973 and is inspired in Physics. It treats all computation as the result of \textit{actors} exchanging \textit{messages} between themselves.

An actor is an object with an address, code and storage. They communicate with each other by sending messages. For this, each of them has a \textit{mailbox}, which is basically just a queue for the incoming messages. Upon receiving a message, an actor will run it's code and can alter it's own storage.

An actor is thus capable of the following actions:
\begin{itemize}
	\item Pulling messages from its mailbox queue.
	\item Deciding what to do with a message, including ignoring it.
	\item Modifying its internal state.
	\item Sending messages to other actors at addresses that it knows about.
	\item Creating new actors.
\end{itemize}

There is no assumed sequence to the above actions and they can be performed in parallel. Messages are asynchronous and thus may arrive in any order.

\section{Specification}
In this section we will delve into the specification of the Enso virtual machine. Enso is heavily modeled after the actor model, with the main difference being that Enso does not support concurrency, although concurrency could in theory be added.

Just like in the actor model, in Enso everything is either an actor or a message.

An actor is an entity composed of the following components:

\begin{itemize}
	\item \textit{ID:} A unique identifier of the actor.
	\item \textit{Code:} A block of code containing functions that can be called by other actors.
	\item \textit{Storage:} A data structure containing arbitrary information and that can be read and modified by the code.
\end{itemize}

Using Enso, the state of the blockchain will be simply the set of all actors. So, all of the application logic and data will be coded into the actors.

Note that actors in Enso do not have a mailbox. That is because, since no concurrency is needed, there is only a single \textit{global message queue} that orders all messages sent by all actors.

Messages, put simply, are asynchronous function calls from one actor to another. By asynchronous we mean that when an actor sends a message, the actor will not be blocked while it waits for the message to be processed. Instead it will keep working and accepting other messages. This asynchronicity is necessary because we want the virtual machine to keep running even if a function call fails.

A message takes the form \texttt{(ID\_to, function\_call, parameters)}, where:

\begin{itemize}
	\item \texttt{ID\_to:} The ID of the actor that will receive the message.
	\item \texttt{function\_call:} The name of the function that will be called.
	\item \texttt{parameters:} A list of parameters that will be passed to the function.
\end{itemize}

Now that we know that the state is composed of actors, and that actors communicate between themselves with messages, we can state which actions are available to an actor in Enso:

\begin{itemize}
	\item Receive messages from the global message queue.
	\item Processing messages by calling the corresponding internal functions.
	\item Modifying its own ID, code or storage.
	\item Send messages to other actors at addresses that it knows about.
	\item Read storage of other actors at addresses that it knows about.
	\item Create new actors.
\end{itemize}

Contrary to the actor model, actors in Enso are able to read each others storage. The blockchain state is public and accessible to anyone, so it only makes sense to allow actors to also read the entire state.

The only component of Enso that remains to be described is the global message queue, which is just a simple \textit{first in first out} queue for messages.

When an message is processed, it is taken out of the queue and the function \texttt{function\_call} of the actor \texttt{ID\_to} is called with the input parameters \texttt{parameters}. If, for some reason, the function cannot be called, for example if the actor doesn't exist or there is no function with that name, the message is simply ignored.

\section{Implementation}
The objective of the last section was to describe the components of the Enso virtual machine and its operation in an abstract form. In this section we want to give some details on how Enso can be implemented in practice.

First, even though both the application data and logic will coexist in the state, in most cases there will be a clear separation between both. By this we mean that, when developing a new blockchain with Enso, it is useful to divide the actors into two categories:

\begin{itemize}
	\item \textbf{Kernel actors:} All actors created at the genesis block and during updates to the blockchain. These are actors that define the rules for how the runtime works and are unique actors, meaning that there is only one instance of each actor type.
	\item \textbf{User actors:} All actors created by the users. There is a predetermined list of actor templates, from which users can create new user actors.
\end{itemize}

The idea is that the entire application logic will be coded into the kernel actors, for example we may have one actor to handle the incoming extrinsics, another that implements some governance method, etc. While the user actors will be regular actors like accounts or smart contracts, effectively representing the application data.

Second, an extrinsic is just a message that is sent by an user, or by some other external source, instead of by an actor. Otherwise, an extrinsic should be able to be processed by an actor just like a message.

Third, a block is just an ordered list of extrinsics. When the virtual machine receives a new block, it just needs to process those extrinsics in order and in the end it will have reached a new state.

So, to process a new block, a node follows these rules:

\begin{enumerate}
	\item Receive an ordered list of extrinsics.
	\item Add all the extrinsics, in order, to the global message queue.
	\item Take the top message and send it to the corresponding actor.
	\item The receiving actor then may create more messages, which will also be added to the queue.
	\item Keep repeating steps 3 and 4 until there are no more messages left
	\item When there are no more messages to be processed, the state transition has finished.
\end{enumerate}

\section{Conclusion}
This paper analyzes the organization of a blockchain's code by thinking of it as a distributed virtual machine. We then argue that a distributed virtual machine needs to act analogously to a stored-program computer, which implies that the application logic needs to be coded into the state and that the state transition function needs to be a general-purpose virtual machine. Using this method it is possible to have blockchains that are simpler and faster to develop and operate, and that require far fewer forks.

\bibliographystyle{IEEEtran}
\bibliography{Enso.bib}

\end{document}